\documentstyle[preprint,eqsecnum,aps,epsf]{revtex}

\begin{document}


\title{Nucleon spin-flavor structure\\  
in\\
SU(3) breaking chiral quark model\\}
\author{X. Song, J. S. McCarthy and H. J. Weber\\
{\it Institute of Nuclear and Particle Physics,}\\
{\it Department of Physics, University of Virginia,}\\
{\it Charlottesville, VA 22901, USA.}}
\maketitle
\begin{abstract}
The SU(3) symmetric chiral quark model, which describes 
interactions between quarks, gluons and the Goldstone
bosons, explains reasonably well many aspects of the
flavor and spin structure of the proton, except for the 
values of $f_3/f_8$ and $\Delta_3/\Delta_8$. Introducing 
the SU(3)-breaking effect suggested by the mass difference 
between the strange and nonstrange quarks, we find that 
this discrepancy can be removed and better overall 
agreement obtained.
\end{abstract}


\widetext

\section{Introduction}

One of the important goals in high energy physics 
is to reveal the internal structure of the nucleon.
This includes the study of the flavor and spin contents  
of the quark and gluon constituents in the nucleon 
and how these are related to the nucleon 
properties: spin, magnetic moment, elastic form 
factors, and deep inelastic structure functions. 
In the late 1980s, the polarized deep inelastic 
lepton nucleon scattering experiments \cite{1} 
surprisingly indicated that only a small portion 
of the proton spin is carried by the quark and 
antiquarks, and a significant negative strange quark 
polarization in the proton sea. Since then, a 
tremendous effort has been made for solving this 
puzzle both theoretically and experimentally (recent 
review see \cite{2,3,4,5}). According to the most recent 
result \cite{6,7}, the quarks contribute about 
one third of proton's spin, which is only one half
of the spin expected from the hyperon decay data
($\Delta\Sigma\simeq 0.6$) and the strange quark 
polarization is about $-0.10$, which deviates 
significantly from the naive quark model expectation. 
On the other hand, the baryon magnetic 
moments can be reasonably well described by the 
spin-flavor structure in the nonrelativistic constituent 
quark model. 

Most recently, the New Muon Collaboration (NMC)
experiments \cite{8} have shown that the Gottfried sum 
rule \cite{9} is violated, which indicates that the 
$\bar d$ density is larger than the $\bar u$ density 
in the nucleon sea. This asymmetry has been confirmed 
by the NA51 Collaboration experiment \cite{10}, which
shows that ${\bar u}/\bar d\simeq 0.51$ at $x=0.18$. 
From the perturbative QCD motivated quark model of the 
nucleon, the density of $\bar u$ would be almost the 
same as that of $\bar d$ if the sea quark pairs are 
produced by the flavor-independent gluons ($\bar s$ 
could be different because of $m_s>>m_{u,d}$). 

Many theoretical works, trying to solve these puzzles, 
have been published. Among these, the application of 
the chiral quark model, suggested by Eichten, Hinchliffe, 
and Quigg \cite{11}, and then extended by Cheng and Li 
\cite{12}, seems to be more promising. The chiral quark 
model was originated by Weinberg \cite{13} and 
then developed by Manohar and Georgi \cite{14}. In this 
model, they introduced an effective Lagrangian for quarks, 
gluons and Goldstone bosons in the region between the chiral 
symmetry-breaking scale ($\Lambda_{\chi SB}\simeq 1$ GeV) 
and the confinement scale ($\Lambda_{QCD}\simeq 0.1-0.3$ GeV). 
The great success of the constituent quark model in low 
energy hadron physics can be well understood in this 
framework. In the chiral quark model the effective 
strong coupling constant $\alpha_s$ could be as small 
as 0.2-0.3, which implies that the hadrons can be 
treated as weakly bound states of effective constituent 
quarks. The model gave a correct value for
$(G_A/G_V)_{n\rightarrow p}=(5/3)\cdot g_A\simeq 1.25$, 
with $g_A\simeq 0.75$, and a fairly good prediction for
baryon magnetic moments. 
 
The extended description given in \cite{12}
can solve many puzzles related to the proton flavor and 
spin structures: a significant strange quark presence 
in the nucleon indicated in the low energy pion nucleon 
sigma term $\sigma_{\pi N}$, the asymmetry between 
$\bar u$ and $\bar d$ densities, the total net quark spin
$\Delta\Sigma\simeq 1/3$ and nonzero negative strange 
polarization $\Delta s\simeq -0.10$. However, the SU(3) 
symmetry description yields $f_3/f_8$=1/3 
and $\Delta_3/\Delta_8$=5/3 (the definitions of $f_{3,8}$ 
and $\Delta_{3,8}$ are given in sec. II; also see \cite{12}), 
which are inconsistent with the experimental values. In 
this paper, we introduce an SU(3) symmetry-breaking effect
that arises from the mass difference between the strange 
and nonstrange quarks, which results in a suppressed 
amplitude for producing the "kaons". The result shows that 
not only can the above discrepancy can be removed but also 
better agreement between some other theoretical predictions 
and experimental results is obtained. The $\eta$ meson is 
not included in the SU(3) breaking here despite its strange 
quark contents because it is not well established as a
Goldstone boson. An investigation in this direction is 
underway and the result will be presented elsewhere. 
Our limited goal in this work is to look at the SU(3) 
breaking by first introducing the suppression of kaon 
fluctuations. 

\section{SU(3) symmetry breaking}

In the scale range between $\Lambda_{\chi SB}$ and 
$\Lambda_{QCD}$ in the chiral quark model, the relevant
degrees of freedom are the quasiparticles of quarks,
gluons and the Goldstone bosons associated with the 
spontaneous breaking of the $SU(3)\times SU(3)$ chiral
symmetry. In this quasiparticle description, the 
effective gluon coupling is small and the important 
interaction is taken to be the coupling among quarks 
and Goldstone bosons, which may be treated as an 
excitation of $q\bar q$ pair produced in the interaction 
between the constituent quark and the quark condensate. 
Note that these Goldstone bosons can be $identified$, $in$
$quantum$ $numbers$, $with$ $the$ $usual$ $pseudoscalar$ 
$mesons$, $but$ $they$ propagate inside the nucleon and 
$are$ $not$ $free$ $on$-$shell$ $mesons$.

The sea quark-antiquark pairs could also be created by 
the gluons. Since the gluon is flavor-independent,
a valence quark cannot change its flavor by emitting a
gluon. Also the spin cannot be changed due to the vector
coupling nature between the quarks and gluons. On the
other hand, emitting a Goldstone boson a valence quark 
could change its flavor and certainly change its spin 
because of pseudoscalar coupling between the quarks and 
Goldstone bosons. The presence of Goldstone bosons in 
the nucleon cause quite a different sea quark flavor-spin 
content from that given by emitting gluons from the quarks. 
Hence the chiral quark model may provide a better 
understanding to the above puzzles.

The effective Lagrangian describing 
interaction between quarks and Goldstone bosons can be
written \cite{12}
\smallskip
$${\it L}_I=g_8{\bar q}{\hat\phi}q
+{\sqrt {1\over 3}}g_0{\bar q}\eta'q ~,\qquad\quad 
q=\pmatrix{ u\cr d\cr s\cr } ~,
\eqno (2.1)$$
where
$${\hat \phi}=\pmatrix{
{1\over {\sqrt 2}}{\pi}^o+{1\over {\sqrt 6}}{\eta}^o & {\pi}^+ & K^+\cr
{\pi}^-& -{1\over {\sqrt 2}}{\pi}^o+{1\over {\sqrt 6}}{\eta}^o & K^o\cr
K^-& {\bar {K^o}}& -{2\over {\sqrt 6}}{\eta}^o\cr }
\eqno (2.2)$$
and $\lambda_i$ (i=1,2,...,8) are the Gell-Mann matrices.
If the singlet Yukawa coupling is equal to zero, $g_0=0$, 
the quark sea created by emitting $0^-$ meson octet would 
contain more $\bar d$ quarks than $\bar u$ quarks (and 
less $\bar s$ quarks). The resulting $\bar u$-$\bar d$ 
asymmetry seems to be consistent with $\bar u$-$\bar d<0$ 
indicated by the NMC data which show a significant violation
of the Gottfried sum rule. However, if the singlet is as 
important as the octet in the quark meson interactions 
and its Yukawa coupling is equal to the octet coupling, 
$g_0=g_8$, then one the flavor asymmetry in the sea 
disappears and the numbers of $u\bar u$, $d\bar d$ and 
$s\bar s$ are equal. Cheng and Li suggested an unequal 
singlet and octet coupling, $g_0/g_8\equiv\zeta \neq 1$
[see discussion below eq (2.12)]. 
Taking $a=|g_8|^2=0.1$ and $\zeta=-1.2$, they obtained
$$\bar u/\bar d=0.53\quad (expt.: 0.51\pm 0.04\pm 0.05),
\eqno (2.3)$$
$$I_G=\int_0^1dx{{F_2^p(x)-F_2^n(x)}\over x}=0.236\quad  
(expt.: 0.235\pm 0.026),
\eqno (2.4)$$
$$f_s\equiv {{2\bar s}\over {3+2({\bar u}+{\bar d}
+{\bar s})}}=0.19 \quad (expt.: 0.18\pm 0.03),
\eqno (2.5)$$
where $f_q=(q+{\bar q})/\sum (q+{\bar q})$ ($q=u,d,s$).
Using the same parameters, the quark spin polarizations are
also consistent with the data (see \cite{12}).

Defining $f_3=f_u-f_d$, $f_8=f_u+f_d-2f_s$, 
$\Delta_3=\Delta u-\Delta d$, and 
$\Delta_8=\Delta u+\Delta d-2\Delta s$, the SU(3) symmetry 
description yields
$$f_3/f_8=1/3\quad (expt.: 0.23),\qquad 
\Delta_3/\Delta_8=5/3\quad (expt.: 2.1)
\eqno (2.6)$$
In Ref.\cite{12}, Cheng and Li suggested that this 
inconsistency between theoretical prediction and 
data could be attributed to some SU(3)-breaking effects. 

We assume that the breaking of SU(3)-flavor symmetry 
arises from a mass difference between the strange
and nonstrange light quarks. Since the $m_s>m_{u,d}$
the breaking would cause a suppressed amplitude,
and thus a smaller probability, for a $u$ quark to
fluctuate into a $K^+=(u\bar s)$ plus a strange quark
than fluctuate into a $pion$ and a nonstrange light quark.
Defining $\Psi(u\rightarrow \pi^+d)$ as the probability 
amplitude
of a $\pi^+$ meson emission from a $u$ quark, etc., we
have
$$|\Psi(u\rightarrow \pi^+d)|^2=
|\Psi(d\rightarrow \pi^-u)|^2=|g_8|^2\equiv a,\quad etc.
\eqno (2.7)$$
for pion emission, and
$$|\Psi(u\rightarrow K^+s)|^2=|\Psi(d\rightarrow K^0s)|^2
\equiv\epsilon a
\eqno (2.8)$$
for kaon emission. The new parameter $\epsilon$ denotes 
the ratio of the probability of emitting a kaon to that 
of a pion from the quarks, and we expect $0<\epsilon\leq 
1$. In the following, we will show that a reasonable 
value $\epsilon\simeq 0.5-0.6$ gives a good fit to the 
data.

It is easy to see that the nonstrange quark numbers, thus
${\bar d}-\bar u$ and $\bar d/\bar u$ would not be affected 
by the SU(3)-breaking effect arising from suppression of 
kaon production, but the strange quark number $\bar s$ and 
$f_s$ would be reduced. A straightforward calculation yields 
the results
$${\bar u}={a\over 3}(\zeta^2+2\zeta+6)~,
\eqno (2.9)$$
$${\bar d}={a\over 3}(\zeta^2+8)~,
\eqno (2.10)$$
$${\bar s}={a\over 3}(\zeta^2-2\zeta+10)-3a(1-\epsilon)~,
\eqno (2.11)$$
which reduce to the SU(3) results \cite{12} when 
$\epsilon\rightarrow 1$. From eqs. (2.9)-(2.11), we obtain 
$${\bar d}-{\bar u}={{2a}\over 3}(1-\zeta),\qquad
{{\bar u}\over {\bar d}}=1-
2({{1-\zeta}\over {\zeta^2+8}})
\eqno (2.12)$$
since data shows $\bar d>\bar u$ and $a>0$, hence $\zeta<1$. 
From the data, ${\bar u}/{\bar d}\simeq 0.51$, which
leads to $(1-{\zeta})/({\zeta^2}+8)\simeq 0.24$, and 
$\zeta\simeq -1.2$. The explanation for $\zeta\neq 1$ 
in \cite{12} was that the nonplanar contributions \cite{15} 
in the $1/N_c$ expansion break the U(3) symmetry. A study
given in \cite{16} shows that the singlet and nonsinglet 
couplings are renormalized differently in the chiral 
quark model due to they receive different contributions
from the loops of Goldstone bosons. A detail model 
calculation in \cite{16} gave $\zeta\simeq -2$. But it 
still needs further study.

For the spin contents of the proton, the SU(3)-breaking results
are
$$\Delta u={4\over 3}-{a\over 9}(8\zeta^2+37)+{{4a}\over 3}(1-\epsilon)~,
\eqno (2.13)$$
$$\Delta d=-{1\over 3}+{2a\over 9}(\zeta^2-1)-{a\over 3}(1-\epsilon)~, 
\eqno (2.14)$$
$$\Delta s=-a+a(1-\epsilon)~,
\eqno (2.15)$$
One can see that with the SU(3)-breaking effect arising from
the kaon suppression, $\Delta u$ would be $more$ $positive$, 
$\Delta d$ $more$ $negative$ and $\Delta s$ $less$ $negative$. 
Compared to the results without SU(3) breaking,
$${\Delta s}-{(\Delta s)_{SU(3)}=a(1-\epsilon),\qquad
{\Delta\Sigma}-{(\Delta\Sigma)_{SU(3)}}}=2a(1-\epsilon).
\eqno (2.16)$$
Hence a consequence of this breaking is that the strange 
sea polarization is reduced (less negative) and the total 
quark spin would slightly increase.

From eqs. (2.9)$-$(2.11), we have
$$f_3/f_8={1\over 3}({1\over {1+{{4a(1-\epsilon)}\over
{1+4a/3(\zeta-1)}}}}).
\eqno (2.17)$$
For the nonsinglet axial charges, one obtains 
$$\Delta_3={5\over 3}[1-{a\over
3}(2\zeta^2+4+3\epsilon)]~,
\eqno (2.18)$$
$$\Delta_8=1-{a\over
3}(2\zeta^2+10-3\epsilon)~,
\eqno (2.19)$$
and
$$\Delta_3/\Delta_8={5\over 3}
({{1-{a\over 3}(2\zeta^2+7)+a(1-\epsilon)}\over
{1-{a\over 3}(2\zeta^2+7)-a(1-\epsilon)}}).
\eqno (2.20)$$
It is obvious that the correction factors, appearing in 
eqs. (2.17) and (2.20), due to the SU(3)-breaking, are 
in the right direction, i.e., $f_3/f_8$ decreases [provided 
that $1+4a/3(\zeta-1)>0$ and $\epsilon<1$] and 
$\Delta_3/\Delta_8$ increases. The SU(3) results can be 
recovered by taking $\epsilon\rightarrow 1$.

\section{Numerical results and discussion}

To maintain the agreement obtained in the SU(3) symmetry
description, we choose $a=0.1$ and $\zeta=-1.2$ used
in ref.\cite{12} [one can see from (2.12) that $a$ and $\zeta$
are completely determined by fitting data $\bar d/\bar u$ and
$\bar d-\bar u$. Two remarks should be made: First, data $\bar
d-\bar u$ is obtained from the measurement of the Gottfried sum
$I_G$ by assuming there is no charge symmetry breaking in the 
sea \cite{ma92}. Second, ``data'' ${\bar d}/{\bar u}$ is 
measured at only a single $x$ value of about 0.18]. For the
SU(3)-breaking parameter $\epsilon$, we choose $\epsilon=0.5-0.6$, 
which is quite a reasonable value if we assume that $\epsilon$ is 
proportional to the ratio $m_{u,d}/m_s$, where $m_{u,d}$ 
and $m_s$ are the constituent quark masses of nonstrange 
and strange quarks. Having three parameters $a$, $\zeta$ 
and $\epsilon$ in the SU(3)-breaking description, one 
obtains 
$$f_3/f_8=0.26\ \ ({\rm expt.}:\ 0.23),\qquad  
\Delta_3/\Delta_8=1.94\ \ ({\rm expt.:}\ 2.10)~;
\eqno (3.1)$$
the theoretical predictions are now much closer to the 
data than those from SU(3) description and the inconsistency 
shown in \cite{12} is removed. The results for the quark 
flavor and spin contents in the proton are listed in 
Tables I and II respectively. Comparison with Cheng-Li's
SU(3)-symmetry prediction and data are shown. For comparison, 
the experimental results from the analysis given by 
Ellis and Karliner \cite{18} are also shown.

Several remarks are in order.

(1) According to the analysis given in \cite{fd96,ratcliffe96}, 
the hyperon $\beta$ decay data can be well accommodated within 
the framework of Cabbibo's SU(3) symmetry description. For 
example, Ref. \cite{fd96} shows that the use of SU(3) symmetry 
with a small SU(3) breaking proportional to the mass difference 
between strange and nonstrange quarks allows a very satisfactory 
description of the hyperon $\beta$ decay data and leaves little 
room for any further SU(3)-breaking contributions. Similar 
conclusion has been reached in \cite{ratcliffe96}. Hence as a
good approximation, one can write 
$$\Delta_3=\Delta u-\Delta d=F+D, \qquad
\Delta_8=\Delta u+\Delta d-2\Delta s=3F-D~.
\eqno (3.2)$$
From the spin contents shown in Table II, we can calculate 
$F/D$ and other weak axial couplings. The results are listed 
in Table III. It shows that our description gives better 
agreement with the hyperon $\beta$ decay data \cite{21,22} as 
well.

(2) The first moments of $g_1^p$ and $g_1^n$ including QCD
corrections can be written as 
$$I^p=\int_0^1dxg_1^p(x)={{C_{NS}}\over 6}\Delta u+{1\over
{18}}(2C_S-C_{NS})\Delta\Sigma~,
\eqno (3.3)$$
$$I^d=\int_0^1dxg_1^d(x)=\eta[-{{C_{NS}}\over 6}\Delta s+
{1\over {18}}(4C_S+C_{NS})\Delta\Sigma]~,
\eqno (3.4)$$
where $\eta=0.4565$ and $C_S(Q^2)$, $C_{NS}(Q^2)$
are the QCD radiative correction factors given in Ref.\cite{23}.
Taking $\alpha_s=0.35$ at $Q^2=3.0$ GeV$^2$ and using the 
spin contents in Table II, the first moments $I^p$ and $I^d$ 
are evaluated and listed in Table IV. 

One can see that the SU(3)-breaking results are also better 
than those without SU(3)-breaking except for the moment of 
$g_1^d$. Our prediction of $I^d$ is higher than both Spin 
Muon Collaboration (SMC) and E143 data. In addition, our $I^p$ 
value is closer to the SMC data, while the SU(3) symmetry 
prediction is closer to the E143 data.

(3) For comparison, we also evaluated a quantity defined as
$$<A_1^p>=2<x>{{\Sigma e_q^2\Delta q}\over {\Sigma e_q^2q}}~,
\eqno (3.5)$$
which is a crude approximation of the asymmetry $A_1^p$ 
measured in deep inelastic lepton proton scattering, 
where $<x>$ is the average value of the Bjorken variable $x$
and can be taken as $0.5-0.7$.
Taking the $q$'s from eqs. (2.9)-(2.11) and $\Delta q$'s
from eqs. (2.13)- (2.15), and using $\alpha_s=0.35$ at 
$Q^2=3.0$ GeV$^2$, we obtain
$$<A_1^p>=0.24-0.34,\ \ (\epsilon=0.5),\qquad 0.20-0.30\ \ 
(\epsilon=1.0);
\eqno (3.6)$$
the data from E143
$$\int_0^1A_1^p(x)dx=0.40\pm 0.10 
\eqno (3.7)$$
seems to favor the symmetry-breaking description.

(4) We decompose the valence and sea contributions for
the flavor contents in the proton. Neglecting the
antisymmetrization effect of the $u$ and $d$ sea quarks 
with the valence quarks ($u$, $d$ in the nucleon), we
may assume $u_{val}=2$ and $d_{val}=1$, since $s_{val}=0$, 
and obtain
$${u}_{sea}={\bar u}={a\over 3}(\zeta^2+2\zeta+6)~,
\eqno (3.8)$$
$${d}_{sea}={\bar d}={a\over 3}(\zeta^2+8)~,
\eqno (3.9)$$
$${s}_{sea}={\bar s}={a\over 3}(\zeta-1)^2+3\epsilon a~, 
\eqno (3.10)$$
here the equality of the sea quark number and the antiquark 
number is because the sea must be $flavorless$. From eqs.
(3.8)$-$(3.10), the sea not only violates SU(3) flavor symmetry 
but also violates SU(2) symmetry: i.e.,
$${\bar s}<{\bar u}<{\bar d}~, 
\eqno (3.11)$$
However, for a special case $\zeta=1$ and $\epsilon=1$, one 
obtains a complete SU(3) symmetric sea: 
${\bar s}={\bar u}={\bar d}$, which was discussed in
\cite{11}.

(5) For sea quark spin contents, we have to be careful in 
defining the sea quark polarizations. Unlike the equality 
${q}^{sea}={\bar q}$ holds in the unpolarized case, we do 
not have similar equality for sea quark polarization and 
corresponding antiquark polarization in general \cite{note,29}.
As an example, the chiral quark model [SU(3) symmetry or 
SU(3) breaking descriptions] gives that all antiquark 
polarizations are zero 
$${\Delta\bar u}={\Delta\bar d}={\Delta\bar s}=0~.
\eqno (3.12)$$
The smallness of antiquark polarizations was discussed in 
\cite{25} and seem to be consistent with the most recent 
SMC experiment \cite{smc96}. It is obvious from eqs. (2.15) 
and (3.12) that $at$ $least$ for the $strange$ $quark$ sea
polarization the equality $\Delta q_{sea}=\Delta\bar q$
does not hold: i.e., 
$$\Delta s_{sea}\neq \Delta\bar s~.
\eqno (3.13)$$ 
If we neglect the antisymmetrization effect of the $u$ 
and $d$ sea quarks with the valence quarks as before, 
we may assume $\Delta u_v={4\over 3}$ and 
$\Delta d_v=-{1\over 3}$, since no valence strange quarks
$\Delta s_{v}=0$, and obtain 
$${\Delta u}_{sea}=-{a\over 9}(8\zeta^2+37)+{{4a}\over 3}(1-\epsilon)~,
\eqno (3.14)$$
$${\Delta d}_{sea}=+{{2a}\over 9}(\zeta^2-1)-{a\over 3}(1-\epsilon)~,
\eqno (3.15)$$
$${\Delta s}_{sea}=-a\epsilon~,
\eqno (3.16)$$
For the SU(3) symmetric case in \cite{12}, one has ($\epsilon=1$) 
$$ {\Delta u}_{sea}<0, \qquad {\Delta d}_{sea}>0, \qquad 
{\Delta s}_{sea}<0~. 
\eqno (3.17)$$
It implies that the sea quark of each flavor is polarized 
in the direction opposite to the valence quark of the same
flavor. However, in the SU(3) breaking-description, 
${\Delta d}_{sea}$ could be negative if 
$1-\epsilon >{2\over 3}(\zeta^2-1)\simeq 0.3$, or 
$\epsilon < 0.7$. In this case, all sea quarks are spinning 
in the opposite direction with respect to the proton spin.
This includes the cases ($\epsilon=0.5$ or 0.6) discussed 
in this work. A set of negative flavor asymmetric sea 
polarizations ($|\Delta s|<|\Delta u|<|\Delta d|$
and $\Delta q<0$ for $q=u,d,s$) has been used in \cite{27}.
The result shows that a set of valence quark helicity 
distributions, given by the c.m. bag model, and a set of 
negative sea helicity distributions can well describe the 
spin dependent structure functions $g_1^p(x)$, $g_1^n(x)$ 
and $g_1^d(x)$ measured in DIS.

(6) If the gluon axial-anomaly contribution \cite{28} is 
taken into account (in the chiral-invariant factorization 
scheme), then one should use 
$\Delta\tilde q=\Delta q-(\alpha_s/{2\pi})\Delta G$,
not the $\Delta q$, to compare with the DIS data.
Taking $(\alpha_s/{2\pi})\Delta G\simeq 0.04$ 
(this implies $\Delta G\simeq 0.7$ at $\alpha_s=0.35$), 
one has (for $\epsilon=0.6$)
$$\Delta\tilde u=0.81,\quad \Delta\tilde d=-0.38,\quad
\Delta\tilde s=-0.10,\quad \Delta\tilde\Sigma=0.33
\eqno (3.18)$$
and
$$I^p=0.125,\quad I^n=-0.033,\quad I^d=0.042
\eqno (3.19)$$
which seem to be in better agreement with the DIS data
listed in Table II and Table IV. Note that the prediction 
for hyperon $\beta$ decay constants, $\Delta_3$, $\Delta_8$ 
and thus $\Delta_3/\Delta_8$ listed in Table III are not 
affected. 

Finally, we have
$${{\mu_p}\over {\mu_n}}=(-{{3}\over 2})[1-{{5a}\over
6}{{1-r\epsilon}\over {1-{{2a}\over 3}
({\zeta}^2+{{11}\over 4})+a(1-\epsilon)}}]~,
\eqno (3.20)$$
where $m_u=m_d$ and $r=m_{u,d}/m_s$ are assumed. It is not
necessary to require $r$ to be equal to the suppression 
parameter $\epsilon$. Assuming $r=\epsilon=0.6$ and using 
the numbers given in Table II, one obtains 
${{\mu_p}/{\mu_n}}\simeq 1.40$, which can be compared to
the data $({{\mu_p}/{\mu_n}})_{expt.}=1.46$. A more detail
discussion on the octet baryon magnetic moments will be 
given elsewhere.

\section{Summary}
 
In this paper we introduced an SU(3)-breaking effect 
into the SU(3) symmetric chiral quark model. A breaking 
parameter $\epsilon\equiv |\Psi(u\rightarrow K^+s)|^2/ 
|\Psi(u\rightarrow \pi^+d)|^2<1$ is suggested. The new 
parameter denotes a smaller probability of the kaon 
emission from a quark than that of emitting pions. 
Taking $\epsilon\simeq 0.5-0.6$, the $f_3/f_8$ and 
$\Delta_3/\Delta_8$ values are much closer to the data. 
With the breaking effect, some other theoretical 
predictions are also in better agreement with the 
experiments. The simple model suggested in this work 
does not have power to predict the flavor and spin 
distributions in the nucleon. Furthermore, no 
$Q^2$-dependence can be discussed in this 
simple calculation. However, the success of explaining 
many puzzles by using only a few parameters encourages 
us to present this work and to study it further.


{\bf Acknowledgments}

One of us (X.S.) would like to thank L. F. Li for useful 
comments and thank X. Ji for helpful discussions. The 
authors thank P. K. Kabir for reading the manuscript and 
suggestions. We also thank D. Crabb and O. Rondon for 
providing the new E142 and E143 data. This work was 
supported in part by the U.S. Department of Energy and 
by the Commonwealth of Virginia, the Institute of Nuclear 
and Particle Physics, University of Virginia.

\vfill\eject

\vspace{0.2 cm}

Table I: Quark flavor contents for the proton in the 
chiral quark model ($a=0.1$, $\zeta=-1.2$) with 
($\epsilon=0.5$ and 0.6) and without ($\epsilon=1.0$) 
SU(3) breaking.
$$
\offinterlineskip \tabskip=0pt 
\vbox{ 
\halign to 1.0\hsize 
   {\strut
   \vrule#                         
   \tabskip=0pt plus 30pt
 & \hfil #  \hfil                  
 & \vrule#                         
 & \hfil #  \hfil                  
 & \vrule#                         
 & \hfil #  \hfil                  
 & \vrule#                         
 & \hfil #  \hfil                  
 & \vrule#                         
 & \hfil #  \hfil                  
 & \vrule#                         
 & \hfil #  \hfil\quad             
   \tabskip=0pt                    %
 & \vrule#                         
   \cr                             
\noalign{\hrule}                   

&Quantity  && Data && $\epsilon$=0.5 &&  $\epsilon$=0.6 && $\epsilon$=1.0 
\cr 
\noalign{\hrule}
& ${\bar u}/{\bar d}$ && $0.51\pm 0.09$ &&  0.53  && 0.53 && 0.53  \cr
& $I_G$ && $0.235\pm 0.026$ && 0.236 && 0.236 &&0.236 \cr
& $f_u$ && $-$  && 0.51  && 0.50 && 0.48  \cr
& $f_d$ && $-$ && 0.35  && 0.35 && 0.33   \cr
& $f_s$ && $0.18\pm 0.03$ && 0.15 && 0.15  && 0.19  \cr
& $f_3/f_8$ && $0.23\pm 0.05$  && 0.26&&0.27 && 0.33 \cr
\noalign{\hrule}
}}$$
\vfill\eject

Table II: Quark spin contents for the proton in the 
chiral quark model ($a=0.1$, $\zeta=-1.2$) with 
($\epsilon=0.5$, 0.6) and without ($\epsilon=1.0$)
SU(3) breaking.
$$
\offinterlineskip \tabskip=0pt 
\vbox{ 
\halign to 1.0\hsize 
   {\strut
   \vrule#                         
   \tabskip=0pt plus 30pt
 & \hfil #  \hfil                  
 & \vrule#                         
 & \hfil #  \hfil                  
 & \vrule#                         
 & \hfil #  \hfil                  
 & \vrule#                         
 & \hfil #  \hfil                  
 & \vrule#                         
 & \hfil #  \hfil                  
 & \vrule#                         
 & \hfil #  \hfil\quad             
   \tabskip=0pt                    %
 & \vrule#                         
   \cr                             
\noalign{\hrule}                   
&Quantity  && Data && $\epsilon$=0.5 &&  $\epsilon$=0.6 && $\epsilon$=1.0 \cr 
\noalign{\hrule}
&$\Delta u$ && $0.84\pm 0.05$\ (E143) && 0.86 &&0.85   && 0.79  \cr
&         && 0.83$\pm$ 0.05 \ (SMC) &&      &&       &&       \cr
&         && 0.85$\pm$ 0.03 \ (E-K) &&      &&       &&       \cr
&$\Delta d$ && $-0.43\pm$ 0.05\ (E143)&& $-$0.34&& $-$0.34 && $-$0.32 \cr  
&         &&  $-0.44\pm$ 0.05\ (SMC)&&      &&       &&       \cr
&         &&  $-0.41\pm$ 0.03\ (E-K)&&      &&       &&       \cr
&$\Delta s$ && $-0.08\pm$ 0.05\ (E143)&& $-$0.05&& $-$0.06 && $-$0.10& \cr 
&         && $-0.09\pm$ 0.05\ (SMC) &&      &&       &&       \cr
&         && $-0.06\pm$ 0.04\ (E142)&&      &&       &&       \cr
&         && $-0.08\pm$ 0.03\ (E-K) &&      &&       &&       \cr
&$\Delta\Sigma$ && 0.30$\pm$ 0.06\ (E143)&& 0.47&& 0.45&& 0.37  \cr  
&             && 0.20$\pm$ 0.11\ (SMC) &&     &&     &&       \cr
&             && 0.39$\pm$ 0.10\ (E142)&&     &&     &&       \cr
&             && 0.37$\pm$ 0.07\ (E-K) &&     &&     &&       \cr
\noalign{\hrule}
}}$$
\vfill\eject

Table III: Hyperon beta-decay constants in the chiral quark 
model ($a=0.1$, $\zeta=-1.2$) with ($\epsilon=0.5$ and 0.6) 
and without ($\epsilon=1.0$) SU(3) breaking.
$$
\offinterlineskip \tabskip=0pt 
\vbox{ 
\halign to 1.0\hsize 
   {\strut
   \vrule#                         
   \tabskip=0pt plus 30pt
 & \hfil #  \hfil                  
 & \vrule#                         
 & \hfil #  \hfil                  
 & \vrule#                         
 & \hfil #  \hfil                  
 & \vrule#                         
 & \hfil #  \hfil                  
 & \vrule#                         
 & \hfil #  \hfil                  
 & \vrule#                         
 & \hfil #  \hfil\quad             
   \tabskip=0pt                    %
 & \vrule#                         
   \cr                             
\noalign{\hrule}                   
&Quantity  && Data && $\epsilon$=0.5 &&  $\epsilon$=0.6 && $\epsilon$=1.0 \cr 
\noalign{\hrule}
&F$-$D   && 1.2573$\pm$ 0.0028&& 1.20        &&1.19  && 1.11    \cr
&F+D/3 && 0.718$\pm$ 0.015  && 0.70        &&0.70  &&  0.67   \cr
&F$-$D   && $-0.340\pm$ 0.017 && $-$0.29       && $-$0.28&& $-$0.22   \cr
&F$-$D/3 && 0.25$\pm$ 0.05    && 0.21        &&  0.21&&  0.22   \cr
&F/D   && 0.575$\pm$ 0.016  && 0.61        &&0.62  &&  0.67   \cr
&$\Delta_3/\Delta_8$ && 2.09$\pm$ 0.13   && 1.94&& 1.88&& 1.67  \cr
\noalign{\hrule}
}}$$
\vfill\eject

Table IV: The first moments of $g_1^{p,n}$ and $g_1^d$ 
in the chiral quark model ($a=0.1$, $\zeta=-1.2$)
with ($\epsilon=0.5$, 0.6) and without ($\epsilon=1.0$)
SU(3) breaking.
$$
\offinterlineskip \tabskip=0pt 
\vbox{ 
\halign to 1.0\hsize 
   {\strut
   \vrule#                         
   \tabskip=0pt plus 30pt
 & \hfil #  \hfil                  
 & \vrule#                         
 & \hfil #  \hfil                  
 & \vrule#                         
 & \hfil #  \hfil                  
 & \vrule#                         
 & \hfil #  \hfil                  
 & \vrule#                         
 & \hfil #  \hfil                  
 & \vrule#                         
 & \hfil #  \hfil\quad             
   \tabskip=0pt                    %
 & \vrule#                         
   \cr                             
\noalign{\hrule}                   
&Quantity  && Data && $\epsilon$=0.5 &&  $\epsilon$=0.6 && $\epsilon$=1.0 \cr 
\noalign{\hrule}
&$I^p$  && 0.136$\pm$ 0.016\ (SMC) && 0.137  &&0.136 &&  0.128  \cr
&       && 0.127$\pm$ 0.011\ (E143)&&        &&      &&         \cr
&$I^n$  && $-0.031\pm$ 0.011\ (E142)&& $-$0.021 && $-$0.022 && $-$0.021\cr
&       && $-0.037\pm$ 0.014\ (E143)&&       &&      &&         \cr
&$I^d$  && 0.034$\pm$ 0.011\ (SMC)  && 0.053 && 0.052&& 0.049   \cr
&       && 0.042$\pm$ 0.005\ (E143) &&       &&      &&         \cr
\noalign{\hrule}
}}$$
\vfill\eject



\baselineskip 16pt
\begin{thebibliography}{99}


\bibitem{1} 
     J.~Ashman et al., {\sl Phys. Lett.} {\bf B206}, 364 (1988); 
Nucl. Phys. {\bf B328}, 1 (1989) 

\bibitem{2}
     R.~L.~Jaffe, {\sl Physics Today}, No. 9, 24 (1995).

\bibitem{3}
     M.~Anselmino, A.~Efremov and E.~Leader, {\sl Phys. Rep.} 
{\bf 261}, 1 (1995).
    
\bibitem{4}
     F.~E.~Close, {\sl hep-ph/9509251}, 1995 (unpublished).

\bibitem{5}
     H.-Y. Cheng, {\sl Int. J. Mod. Phys.} {\bf A11}, 5109 (1996).

\bibitem{6}
     B.~Adeva {\it et al.} {\sl Phys.Lett.} {\bf B302}, 553 (1993);\ 
     {\bf B320}, 400 (1994)\\
     D.~Adams {\it et al.}, {\sl Phys.Lett.} {\bf B329}, 399 (1994);\
     {\bf B336}, 125 (1994).

\bibitem{7}
     P.~L.~Anthony {\it et al.} {\sl Phys.Rev.Lett.} {\bf 71}, 959 (1993);\
K.~Abe {\it et al.} {\sl ibid.} {\bf 74}, 346 (1995);\
{\bf 75}, 25 (1995).

\bibitem{8}
     New Muon Collaboration, P. Amaudruz {\it et al.}, 
{\sl Phys. Rev. Lett.} {\bf 66}, 2712 (1991;\ M. Arneodo {\it et al.}, 
{\sl Phys. Rev.} {\bf D50}, R1, (1994).

\bibitem{9} 
     K. Gottfried, {\sl Phys. Rev. lett.} {\bf 18}, {1174} (1967).

\bibitem{10} 
     NA51 Collaboration, A. Baldit {\it et al.}, {\sl 
Phys. Lett.} {\bf B332}, {244} (1994).

\bibitem{11} 
     E. J. Eichten, I. Hinchliffe and C. Quigg, {\sl 
Phys. Rev.} {\bf D45} {2269} (1992);\ see also J. D. Bjorken, 
in {\it Elastic and Diffractive Scattering}, Proceedings of the
International Conference, la Biodola, Italy, 1991, edited by 
Cervelli and S. Z. Zucchelli [Nucl. Phys. B, Suppl. (Proc. Suppl.)
{\bf 25B} (1992)]

\bibitem{12} 
     T. P. Cheng and Ling-Fong Li, {\sl Phys. Rev. 
Lett.} {\bf 74} {2872} (1995).

\bibitem{13} 
     S. Weinberg, {\sl Physica} {\bf A96}, {327} (1979).

\bibitem{14} 
     A. Manohar and H. Georgi, {\sl Nucl. Phys.} {\bf B234}, {189} (1984).

\bibitem{15} 
     G, Veneziano, {\sl Nucl. Phys.} {\bf B159}, {213} (1979);
{\bf B117}, {519} (1977).
.
\bibitem{16} 
     X. Li and Y. Liao, {\sl Phys. Lett} {\bf B379} {219} (1996).

\bibitem{ma92} 
     B. Q. Ma, {\sl Phys. Lett} {\bf B274} {111} (1992).

\bibitem{18} 
     J. Ellis and M. Karliner, {\sl Phys. Lett} {\bf B341} {397} (1995).

\bibitem{fd96} 
     X. Song, P. K. Kabir and J. S. McCarthy, {\sl Phys. 
Rev.} {\bf D54} 2108 (1996)

\bibitem{ratcliffe96} 
     P. G. Ratcliffe, {\sl Phys. Lett.} {\bf B365} 383 (1996)

\bibitem{21} 
     Particle Data Group, L. Montanet et al., {\sl Phys. 
Rev.} {\bf D50} {1173} (1994).

\bibitem{22} 
     F. E. Close and R. G. Roberts, {\sl Phys. Lett.} {\bf b316}
{165} (1993).

\bibitem{23}
     S.~A.~Larin, {\sl Phys. Lett.} {\bf B334}, 192 (1994)

\bibitem{note} 
     The equality ${q}_{sea}={\bar q}$ is usually taken as 
the $definition$ of $q_{sea}$, but it is more appropriate 
to say that the equality is required by the $flavorlessness$ 
$of$ $the$ $sea$. For the sea polarization, we do not have 
similar requirement in general. If one $defines$ 
$${\Delta q}_{sea}={\Delta\bar q}$$ it implies 
$(q_{sea})_{\uparrow}-(q_{sea})_{\downarrow}
={\bar q_{\uparrow}}-{\bar q_{\downarrow}}$. 
Considering the equality ${q}_{sea}={\bar q}$, which gives
$(q_{sea})_{\uparrow}+(q_{sea})_{\downarrow}
={\bar q_{\uparrow}}+{\bar q_{\downarrow}}$, we immediately
find that
$$(q_{sea})_{\uparrow}={\bar q_{\uparrow}},\qquad 
(q_{sea})_{\downarrow}={\bar q_{\downarrow}}\qquad
(q=u,d,s,...)~;$$
these are very strong constraints imposed on the sea spin 
components. It may be true for the quark-antiquark pair 
produced from gluons but not necessarily true in general 
case. For example, in the chiral quark model, one has 
$\Delta q_{sea}\neq 0$ and $\Delta\bar q$=0. After 
completion of this paper, we have seen a paper \cite{29}
which reached a similar conclusion in their light-cone 
meson-baryon fluctuation model.

\bibitem{25} 
     T. P. Cheng and Ling-Fong Li, {\sl Phys. Lett.} {\bf B366}
365 (1996).

\bibitem{smc96} 
     B.~Adeva {\it et al.} {\sl Phys.Lett.} {\bf B369}, 93 (1996).

\bibitem{27} 
     X. Song and J. S. McCarthy, {\sl Phys. Rev.} {\bf D49} {3169}
(1994);\ see also X. Song, Proceeding of the Second International 
Symposium on Medium Energy Physics, p.101, (World Scientific 1994).

\bibitem{28} 
     A. V. Efremov and O. V. Teryaev, Dubna report JIN-E2-88-287, 1988
(unpblished); \\
     G. Altarelli and G. Ross, {\sl Phys. Lett.} {\bf B212} 391 (1988);\\
     R. D. Carlitz, J. D. Collins and A. H. Mueller, {\sl Phys. Lett.} 
          {\bf B214} 219 (1988).

\bibitem{29} 
     S. J. Brodsky and B. Q. Ma, {\sl Phys. Lett.} {\bf B381} 317 (1996).

 
\end{thebibliography}
\end{document}